\documentclass[aps,prd,showpacs,nofootinbib,superscriptaddress,preprintnumbers,twocolumn]{revtex4-1}
\usepackage[utf8]{inputenc}
\usepackage{graphicx,amsmath,amssymb,soul,bbm,subcaption,dcolumn}
\usepackage{booktabs, multirow,array,float,enumitem}
\usepackage{caption}
\newcolumntype{C}{>{$}c<{$}}
\AtBeginDocument{
\heavyrulewidth=.08em
\lightrulewidth=.05em
\cmidrulewidth=.03em
\belowrulesep=.65ex
\belowbottomsep=0pt
\aboverulesep=.4ex
\abovetopsep=0pt
\cmidrulesep=\doublerulesep
\cmidrulekern=.5em
\defaultaddspace=.5em
}
\usepackage[dvipsnames]{xcolor}
\usepackage{lmodern,dsfont}
\newcommand{\be}{\begin{eqnarray}}
\newcommand{\ee}{\end{eqnarray}}

\def\MeV {\mathop{\hbox{MeV}}}

\def\Re {\mathop{\hbox{Re}}}
\def\Im {\mathop{\hbox{Im}}}

\usepackage{amsmath} % wonderful math package
\usepackage{dsfont}  % double strike font
\usepackage{bm}      % bold math

\def\beq{\begin{equation}}
\def\eeq{\end{equation}}
\def\beqs#1\eeqs{\beq\begin{split} #1 \end{split}\eeq}

\def\comment#1{}

\usepackage{microtype}

\usepackage[colorlinks=true,backref=false, linktocpage=true,
citecolor=PineGreen,urlcolor=PineGreen,linkcolor=PineGreen,pdfpagemode=UseOutlines]{hyperref}

\hypersetup{%
  bookmarksnumbered=true,
  pdftitle = {},
  pdfsubject = {},
  pdfauthor = {},
  pdfkeywords = {}
}

\begin{document}
\title{Sigma resonance parameters from a $N_f=2$ lattice QCD simulation}

\author{R.\ Molina}
\email{ramope@if.usp.br}
\affiliation{Institute of Physics of the University of S\~ao Paulo, Rua do Mat\~ao, 1371, Butant\~a, S\~ao Paulo, 05508-090, Brazil}

\author{D.\ Guo}
\email{guodehua@gwmail.gwu.edu}
\affiliation{The George Washington University, Washington, DC 20052, USA}

\author{A.\ Alexandru}
\email{aalexan@gwu.edu}
\affiliation{The George Washington University, Washington, DC 20052, USA}
\affiliation{Department of Physics, University of Maryland, College Park, MD 20742, USA}
\affiliation{Albert Einstein Center for Fundamental Physics, Institute for Theoretical Physics, University of Bern, Sidlerstrasse 5, CH-3012 Bern, Switzerland}

\author{M.\ Mai}
\email{maximmai@gwu.edu}
\affiliation{The George Washington University, Washington, DC 20052, USA}

\author{M.\ D\"oring}
\email{doring@gwu.edu}
\affiliation{The George Washington University, Washington, DC 20052, USA}
\affiliation{Thomas Jefferson National Accelerator Facility, Newport News, VA 23606, USA}

%\preprint{JLAB-THY-18-\cor{}{XXXX}}

\begin{abstract}
In this work we present the analysis of the energy spectrum from a recent two-flavor ($N_f=2$) lattice QCD calculation for pion-pion scattering in the scalar, isoscalar channel (the $\sigma$-meson). The lattice simulation was performed for two quark masses corresponding to a pion mass of $315\MeV$ and $227\MeV$. The $\sigma$-meson parameters are extracted using various parametrizations of the scattering amplitude. The results obtained from a chiral unitary parametrization are extrapolated to the physical point and read ${M_\sigma = (440^{+10}_{-16}(50) - i\,240(20)(25))\MeV}$, where the uncertainties in the parentheses denote the stochastic and systematic ones. The behavior of the $\sigma$-meson parameters with increasing pion mass is discussed as well.
\vspace{1cm}\\
{\it XIV International Workshop on Hadron Physics,}\\{\it 18-23 March 2018,}\\{\it Florian\'opolis, SC, Brazil}\end{abstract}

%\pacs
%{
%12.38.Gc, % Lattice QCD calculations
%14.40.-n, %properties of mesons
%13.75.Lb  %meson meson interactions
%}
\maketitle

%%%%%%%%%%%%%%%%%%%%%%%%%%%%%%%%%%%%%%%%%%%%%%%%%%%%%%%%%%%%%
\section{Introduction}

The $\sigma$ is one of the resonances more discussed in the literature. Several phenomenological models have pointed out that it is relevant for nuclear forces. However, its properties have been subject to a long discussion due to a variety of models which tried to determine its mass and width from experimental data, often showing incompatible results.
Even its existence was several times questioned and the Review of Particle Properties did not show it for twenty years. The $\sigma$ particle is nowadays a well-established resonance whose pole properties have been determined more precisely from dispersion relations and coupled-channel approaches which rely upon chiral Lagrangians, analyticity, and unitarity relations~\cite{Pelaez:2015qba}.

Recently, the GWU group has extracted the energy spectrum and phase shifts for pion-pion scattering in the scalar, isoscalar channel, in a $N_f=2$ lattice QCD simulation for two different pion masses, $m_\pi=227$ and $315$ MeV. In this work, we present the analysis of these data
by using several parameterizations and estimate the uncertainties due to the use of one or other. First, we consider a general expansion in an energy-variable conformally  mapping the energy plane to the unit disk, similar to the analysis of Refs.~\cite{Yndurain:2007qm, Caprini:2008fc}. Second, we employ a model based on the chiral unitary approach (UChPT), used, e.g.,
in Refs.~\cite{Hu:2017wli,Hu:2016shf,Guo:2016zos}.  
Subsequently, the UChPT amplitude is extrapolated to the physical point. Our final result, 
based on all lattice data presented here with and without the isovector channel data~\cite{Guo:2016zos}, 
reads $M_\sigma^{\rm phys}=(440^{+10}_{-16}-i\,240_{-20}^{+20})$~MeV and agrees with the result of the  
most recent analysis of experimental data~\cite{Pelaez:2015qba} within the quoted $1\sigma$ region. 
Additionally, the study of the pion mass dependence of resonance mass and coupling to the $\pi\pi$ 
channel is studied in a broader range of $M_\pi$, as it was done in previous works~\cite{Doring:2016bdr,Hanhart:2008mx}.

%%%%%%%%%%%%%%%%%%%%%%%%%%%%%%%%%%%%%%%%%%%%%%%%%%%%%%%%%%%%%

%%%%%%%%%%%%%%%%%%%%%%%%%%%%%%%%%%%%%%%%%%%%%%%%%%%%%%%%%%%%%%%
\section{Analysis of the $\pi\pi$ scattering amplitude}
\label{sec:analysis}

%%%%%%%%%%%%%%%%%%%%%%%%%%%%%%%%%%%%%%%%%%%%%%%%%
\begin{figure*}[t]
 \includegraphics[scale=0.4]{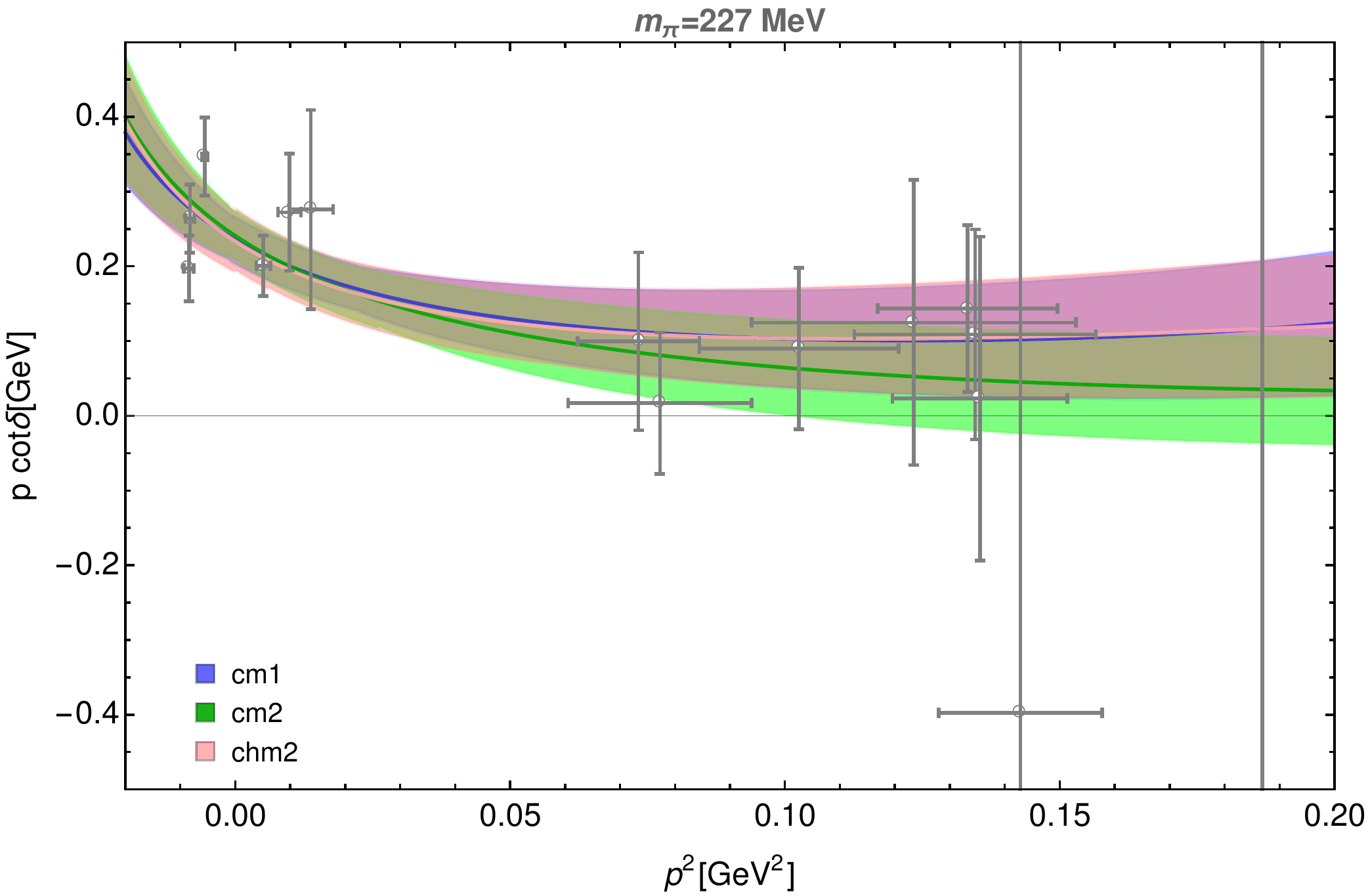}
 \includegraphics[scale=0.4]{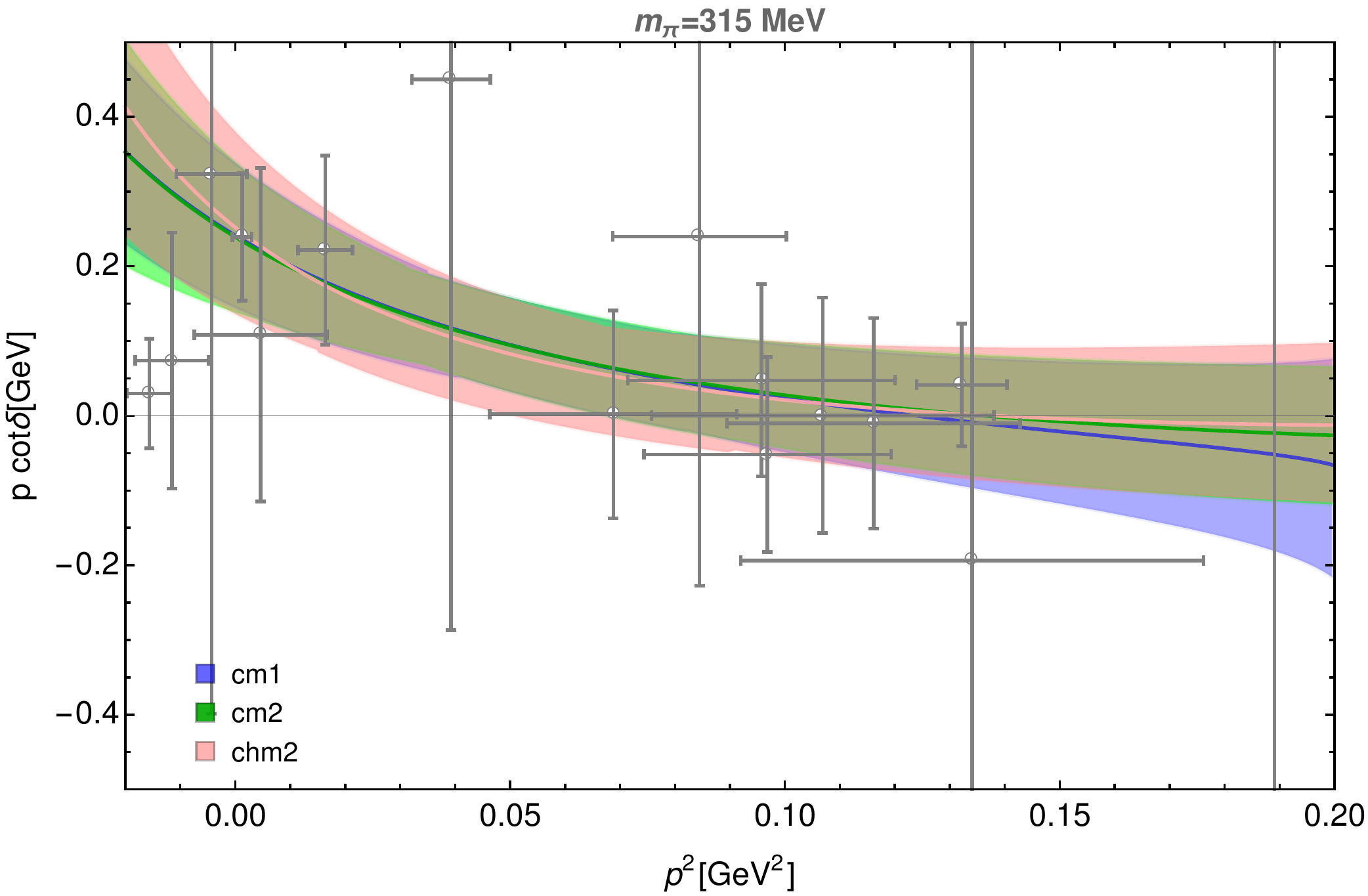}
\caption{Results of the individual fits, for $M_\pi=227$~MeV or $M_\pi=315$~MeV, to the lattice data in terms of $p\,\mathrm{cot}\,\delta$ using the conformal parametrization in two variants ([cm1] and [cm2]), and the chiral unitary approach ([chm2]).
}
\label{fig:conformal}
\end{figure*}
%%%%%%%%%%%%%%%%%%%%%%%%%%%%%%%%%%%%%%%%%%%%%%%%%

The discrete energy spectrum is obtained from the finite volume scattering amplitude,
\begin{align}
\tilde{T}(s)=\frac{1}{K^{-1}(s)-\tilde{G}(s)}\,,
\label{eq:Kmatrix}
\end{align}
which depends explicitly on the form of the $K$-matrix~\cite{Doring:2012eu}.
For boxes with asymmetry $\eta$ in the $z$ direction and in the rest frame,
\begin{align}
\label{eq:gtilde}
&\tilde{G}(s)=G(s)
\\&+
\lim\limits_{q_\mathrm{max}\to \infty}
\left(
\frac{1}{\eta L^3}\sum_{|{\bf q}|<q_{\mathrm{max}}}I(s,|{\bf q}|)
-
\int_{|{\bf q}|<q_{\mathrm{max}}}\frac{d^3q}{(2\pi)^3}I(s,|{\bf q}|)\right)\nonumber
\\
&\text{for~}
I(s,|{\bf q}|)=\frac{\omega_1+\omega_2}{2\omega_1\omega_2}\frac{1}{s-(\omega_1+\omega_2)^2}\,, \nonumber
%&\equiv G(s)+\lim\limits_{q_{\mathrm{max}}\to\infty}\delta G\,.
\end{align}
and $\omega_i=\sqrt{|{\bf q}|^{2}+m_i^2}$, being ${\bf q}$ the momentum in the rest frame, ${\bf q}=2\pi/L\,(n_x,n_y,n_z/\eta)$, and $G(s)$ is two pion loop-function written in the conventional dimensional regularization form \cite{Guo:2018zss}.
The positions of poles in Eq. (\ref{eq:Kmatrix}) determine the discrete energy spectrum on the lattice. The form of the function $K(s)$ is not fixed by unitarity. In this work we use four versions of two different types of the $K$-matrix to gauge the systematic uncertainty tied to a particular choice.
We analyze data from different pion masses and channels individually and also simultaneously. The parameterizations, channels and pion masses employed and number of parameters (pars.) in the fits are:\\
- Individual fits, $m_\pi=227$ or $315$ MeV,
\begin{itemize}
\item[1)] Conformal parameterizations, $\sigma$ (2 pars.)
\item[2)] Chiral parameterization, $\sigma$ and $\rho$ (4 pars.)
\end{itemize}
- Combined fits, for $m_\pi=227$ and $315$ MeV,
\begin{itemize}
\item[3)] Chiral parameterization, $\sigma$ (3 pars.)
\item[4)] Chiral parameterization, $\sigma$ and $\rho$ (4 pars.).
\end{itemize}
For the conformal parameterizations, Eqs. (7) and (8), and (26) and (27) from Ref. \cite{Caprini:2008fc}, are used. These are named [cm1] and [cm2] respectively. The relation of $\psi$ in Ref. \cite{Caprini:2008fc} with the K-matrix in Eq. (\ref{eq:Kmatrix}), is $K^{-1}=-p_{cm}\,\psi/8\pi\sqrt{s} \rho(s)$, with $\rho(s)=\sqrt{1-4 m^2_\pi/s}$. For the chiral parameterizations, we employ the chiral unitary approach described in Ref. \cite{Oller:1998hw}. 
We call them [chm1] and [chm2] for $\sigma$ fits, and $\sigma+\rho$ analyses, respectively.
The data corresponding to the $I=L=1$ $\rho$-channel was analyzed previously in Ref. ~\cite{Guo:2016zos}. The same set of data is used in this work. The above parameterizations and details of the fits are described in detail in Refs. \cite{Guo:2016zos,Guo:2018zss}. Below, we summarize our main results.

%%%%%%%%%%%%%%%%%%%%%%%%%%%%%%%%%%%%%%%%%%%%%%%%%
\begin{figure*}[t]
 \includegraphics[scale=0.4]{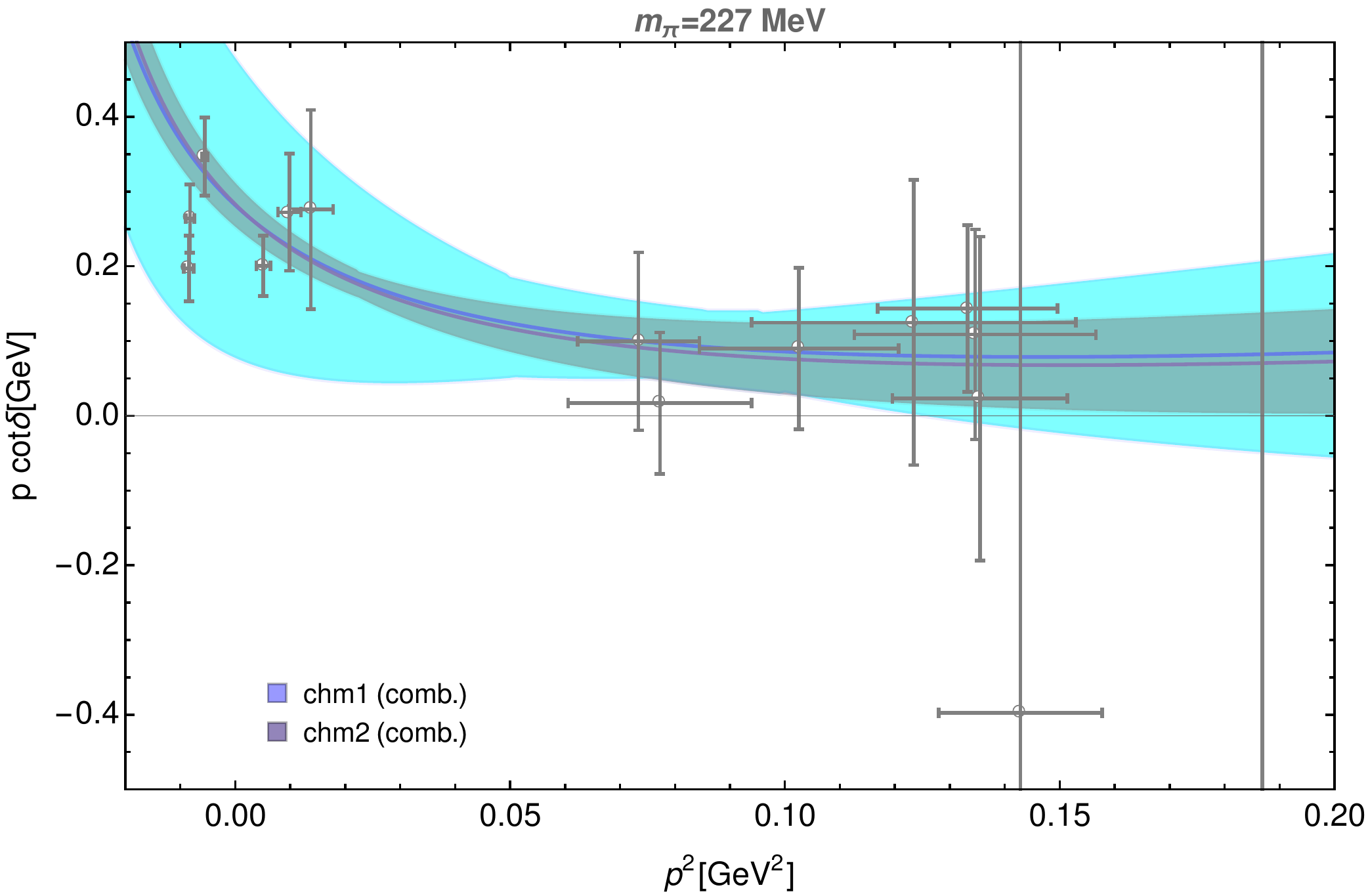}
 \includegraphics[scale=0.42]{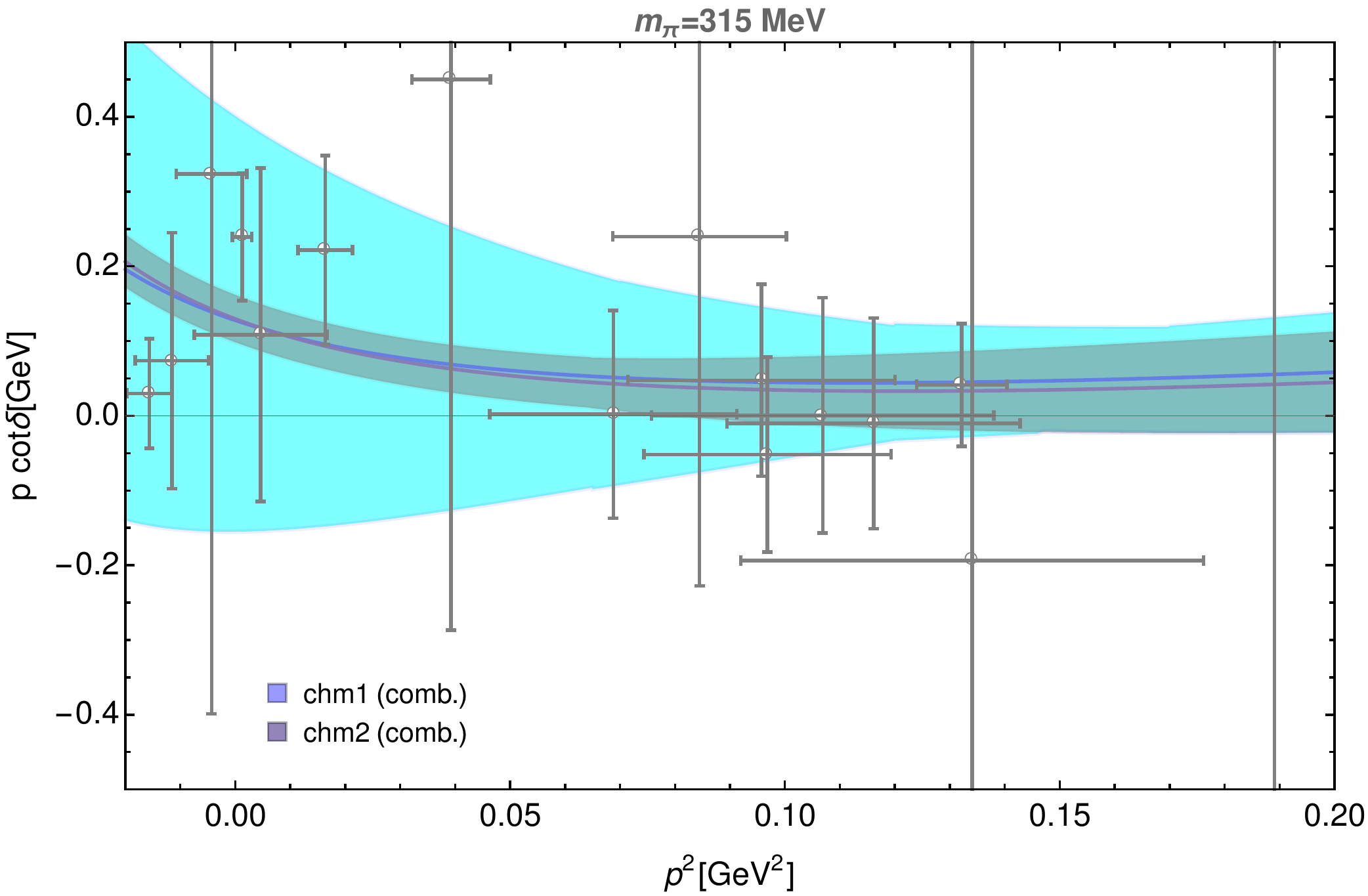}
\caption{Results of the fits to the lattice data using chiral parametrization as described in the main text. The fits are performed simultaneously to the data of both pion masses ($M_\pi=227$ and $315$ MeV), including only the data in the $\sigma$ channel [chm1], or the $\sigma$ and $\rho$ channel [chm2].
}
\label{fig:uchpt}
\end{figure*}
%%%%%%%%%%%%%%%%%%%%%%%%%%%%%%%%%%%%%%%%%%%%%%%%%

\section{Results}

The results of the fits are collected in Table I of Ref. \cite{Guo:2018zss}, which all pass the Pearson's test with the total $\chi^2$ lying inside of the 80\% confidence interval. We discuss first the results from individual fits. 
The phase shifts are depicted in Fig. \ref{fig:conformal}. There is a clear overlap of the fits with the data. Only for the heavy pion mass and energies deep below threshold there is some discrepancy that could be significant. 
Note again that the phase-shifts are not fitted directly, but rather the energies extracted from lattice QCD. There is a good agreement between the phase shifts from both conformal parameterizations used in the $\sigma$ channel [cm1] and [cm2], that
we are interested here, and also with the $\sigma+\rho$ [chm2] chiral individual fits. Note that we used the conformal parameterizations with two parameters. Since the chiral model for $\sigma$ alone has three parameters, uncertainties are larger if we only 
perform individual $\sigma$ fits. Indeed, the chiral parameterization contains LECs which appear in both, $\sigma$ and $\rho$ channels. Thus, the fit of both channels, $\sigma$ and $\rho$, constrains better the LECs used.

We perform an analytical continuation to the complex energy plane. On the second Riemann sheet of this plane we determine the position ($z_0$) and residuum ($g^2$) of the $\sigma$-resonance pole. The results are collected in Table~\ref{tab:poles}. In the three different parameterizations, and for individual fits at $m_\pi=227$ or $315$ MeV, there is a good agreement in the pole position, which lies above threshold, being a resonance in both cases.

\begin{figure}[h!]
\includegraphics[width=1\linewidth]{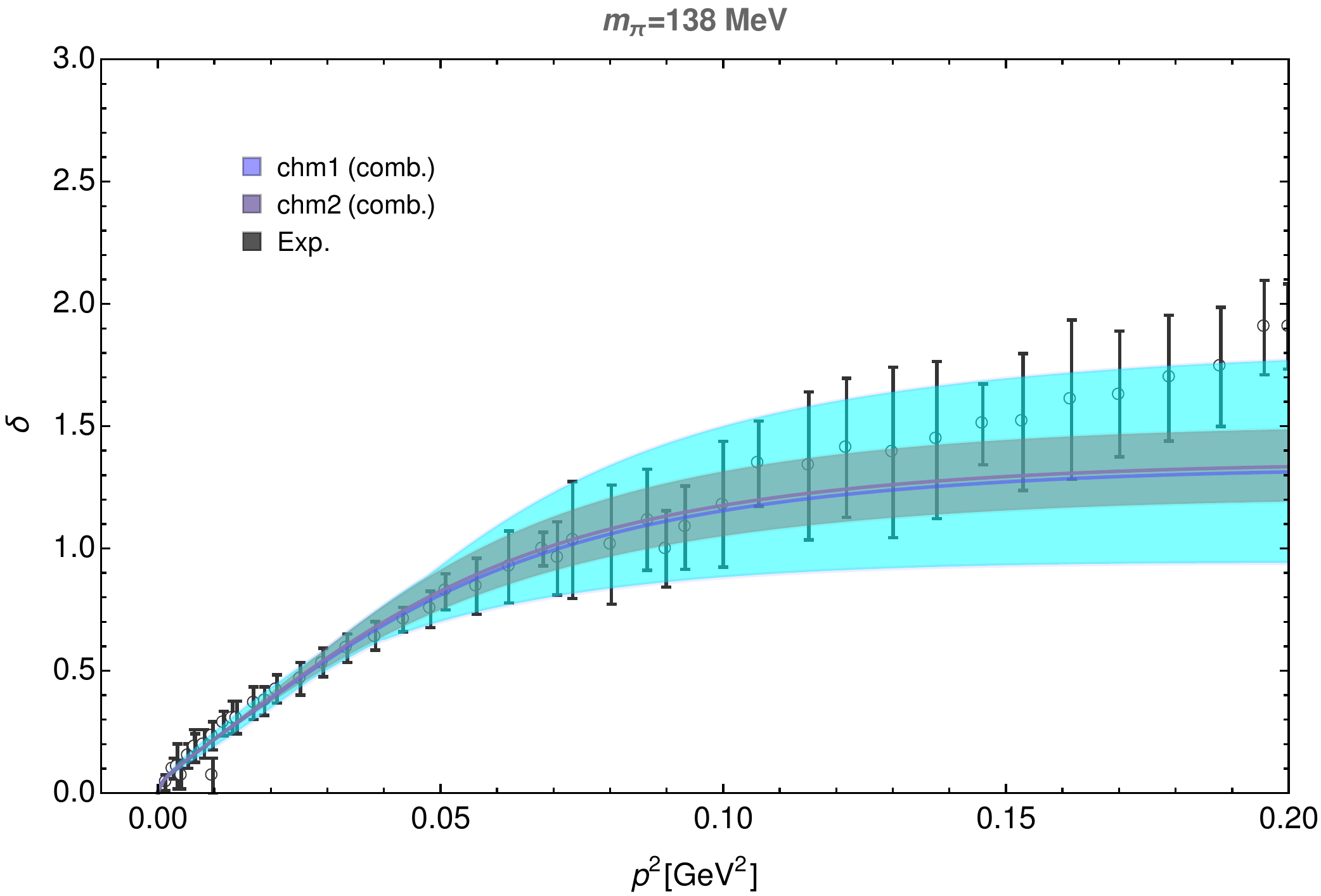}
\caption{
\label{fig:uchpt-expt}
Phase-shift extrapolated to the physical point as a result of the fits to the energy levels in the $\sigma$ (red) and the $\sigma+\rho$ (blue) channels, in comparison with the experimental data, see Ref. \cite{Guo:2018zss}.}
\end{figure}

For combined fits, the corresponding phase-shifts are depicted in Fig.~\ref{fig:uchpt}. As shown in this picture, fitting the $\sigma$ channel alone leads to larger uncertainties in the chiral unitary model, than in combination with the $\rho$ channel. Note that
uncertainites are also larger in this channel for the raw lattice data. The best fit parameters are collected in Table~I of Ref. \cite{Guo:2018zss}.
For the low-energy constants we observe agreement between the individual and combined $\sigma+\rho$ fits results. The extrapolated phase-shifts to the physical point are depicted in Fig.~\ref{fig:uchpt-expt}, showing a good agreement with the experimental data below energies of $950$ MeV. The results for the pole properties are collected in the rows 5-8 of Table~\ref{tab:poles}. 
The $\sigma$ pole position and residuum at the physical mass is compatible in both chiral parameterizations, [chm1] and [chm2]. There is also agreement in the pole properties for the other pion masses. Note that, in the case of the heavy pion mass, the pole position is pushed towards the threshold in the combined fit because of the presence of the light pion mass, becoming a virtual state. The uncertainites for the pole properties at this mass are larger, partly because the state is near the transition to a virtual state, and partly because of the larger uncertainites in the raw lattice data.

%%%%%%%%%%%%%%%%%%%%%%%%%%%%%%%%%%%%%%%%%%%%%%%%%

%%%%%%%%%%%%%%%%%%%%%%%%%%%%%%%%%%%%%%%%%%%%%%%%%
%%%%%%%%%%%%%%%%%%%%%%%%%%%%%%%%%%%%%%%%%%%%%%%%%

%%%%%%%%%%%%%%%%%%%%%%%%%%%%%%%%%%%%%%%%%%%%%%%%%

\begin{table*}[th]
{\renewcommand{\arraystretch}{1.4}
\begin{tabular}{lllllllllllll}
%\toprule
&
&\multicolumn{3}{c}{$M_\pi=138$ MeV}&
&\multicolumn{3}{c}{$M_\pi=227$ MeV}&
&\multicolumn{3}{c}{$M_\pi=315$ MeV}\\
\cline{3-5}
\cline{7-9}
\cline{11-13}
Parametrization~~&Fitted data
&~~$\Re z^*$&~~$-\Im z^*$&~~$g$&
&~~$\Re z^*$&~~$-\Im z^*$	&~~$g$&
&~~$\Re z^*$&~~$-\Im z^*$	&~~$g~~$\\    
\toprule
cm1&$\sigma_{227}$
&~~--&~~--&~~--&
&~~$460_{-60}^{+30}$&~~$180_{-30}^{+30}$&~~$3.2_{-0.1}^{+0.1}$&
&~~--&~~--&~~--\\
cm1&$\sigma_{315}$
&~~--&~~--&~~--&
&~~--&~~--&~~--&
&~~$660^{+50}_{-70}$&~~$150^{+40}_{-50}$&~~$4.0_{-0.2}^{+0.2}$\\
\midrule
cm2&$\sigma_{227}$
&~~--&~~--&~~--&
&~~$475_{-60}^{+30}$&~~$176^{+50}_{-40}$&~~$3.3_{-0.2}^{+0.3}$&
&~~--&~~--&~~--\\
cm2&$\sigma_{315}$
&~~--&~~--&~~--&
&~~--&~~--&~~--&
&~~$660_{-90}^{+50}$&~~$140^{+40}_{-50}$&~~$3.9_{-0.2}^{+0.2}$\\
\midrule
chm1&$\sigma_{227,315}$
&~~$440^{+60}_{-90}$&~~$240^{+20}_{-50}$&~~$3.0^{+0.2}_{-0.6}$&
&~~$490^{+100}_{-70}$&~~$170^{+40}_{-110}$&~~$3.0^{+0.7}_{-0.5}$&
&~~$590^{+130}_{-120}$&~~$80^{+150}_{-80}$&~~$4.0^{+4.0}_{-2.0}$\\
\midrule
chm2&$\sigma_{227}~\rho_{227}$
&~~$430^{+20}_{-30}$&~~$250^{+30}_{-30}$&~~$3.0^{+0.1}_{-0.1}$&
&~~$460^{+30}_{-40}$&~~$160^{+30}_{-30}$&~~$3.0^{+0.1}_{-0.1}$&
&~~$620^{+10}_{-80}$&~~$0^{+60}_{-0}$&~~$3.1^{+6.0}_{-3.0}$\\
chm2&$\sigma_{315}~\rho_{315}$
&~~$460^{+10}_{-15}$&~~$210^{+40}_{-30}$&~~$3.0^{+0.1}_{-0.1}$&
&~~$540^{+30}_{-40}$&~~$150^{+30}_{-30}$&~~$3.1^{+0.1}_{-0.1}$&
&~~$660^{+40}_{-60}$&~~$120^{+40}_{-40}$&~~$3.6^{+0.1}_{-0.1}$\\
chm2&$\sigma_{227,315}~\rho_{227,315}$
&~~$440^{+10}_{-16}$&~~$240^{+20}_{-20}$&~~$3.0_{-0.0}^{+0.0}$&         
&~~$500_{-20}^{+20}$&~~$160_{-15}^{+15}$&~~$3.0^{+0.0}_{-0.1}$&
&~~$600^{+30}_{-40}$&~~$80^{+20}_{-80}$&~~$3.9^{+5.0}_{-0.2}$\\
\toprule     
Ref.~\cite{Pelaez:2015qba}& experimental
&~~$449_{-16}^{+22}$&~~$275_{-12}^{+12}	$&~~$3.5^{+0.3}_{-0.2}$&
&~~--&~~--&~~--&
&~~--&~~--&~~--\\
\hline
\end{tabular}}
\caption{
\label{tab:poles}
Pole positions ($z^*$ in MeV) and corresponding couplings to the $\pi\pi$ channel ($g$ in GeV) from conformal mapping ([cm1] and [cm2]) and chiral unitary approach ([chm1] and [chm2]).}
\end{table*}
%%%%%%%%%%%%%%%%%%%%%%%%%%%%%%%%%%%%%%%%%%%%%%%%%

%%%%%%%%%%%%%%%%%%%%%%%%%%%%
\begin{figure*}[th]
\includegraphics[height=5.cm]{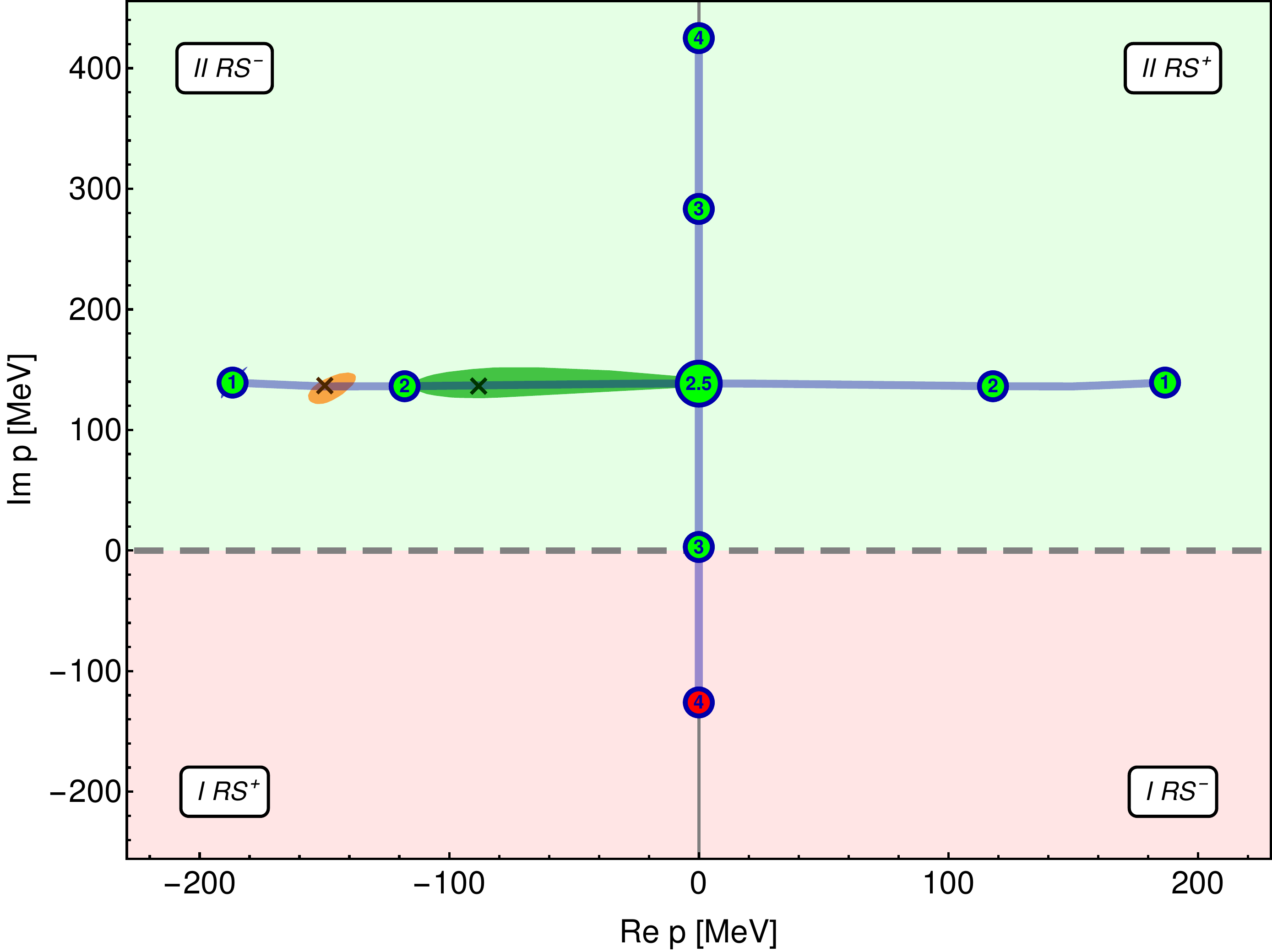}
~~
\includegraphics[height=5.cm]{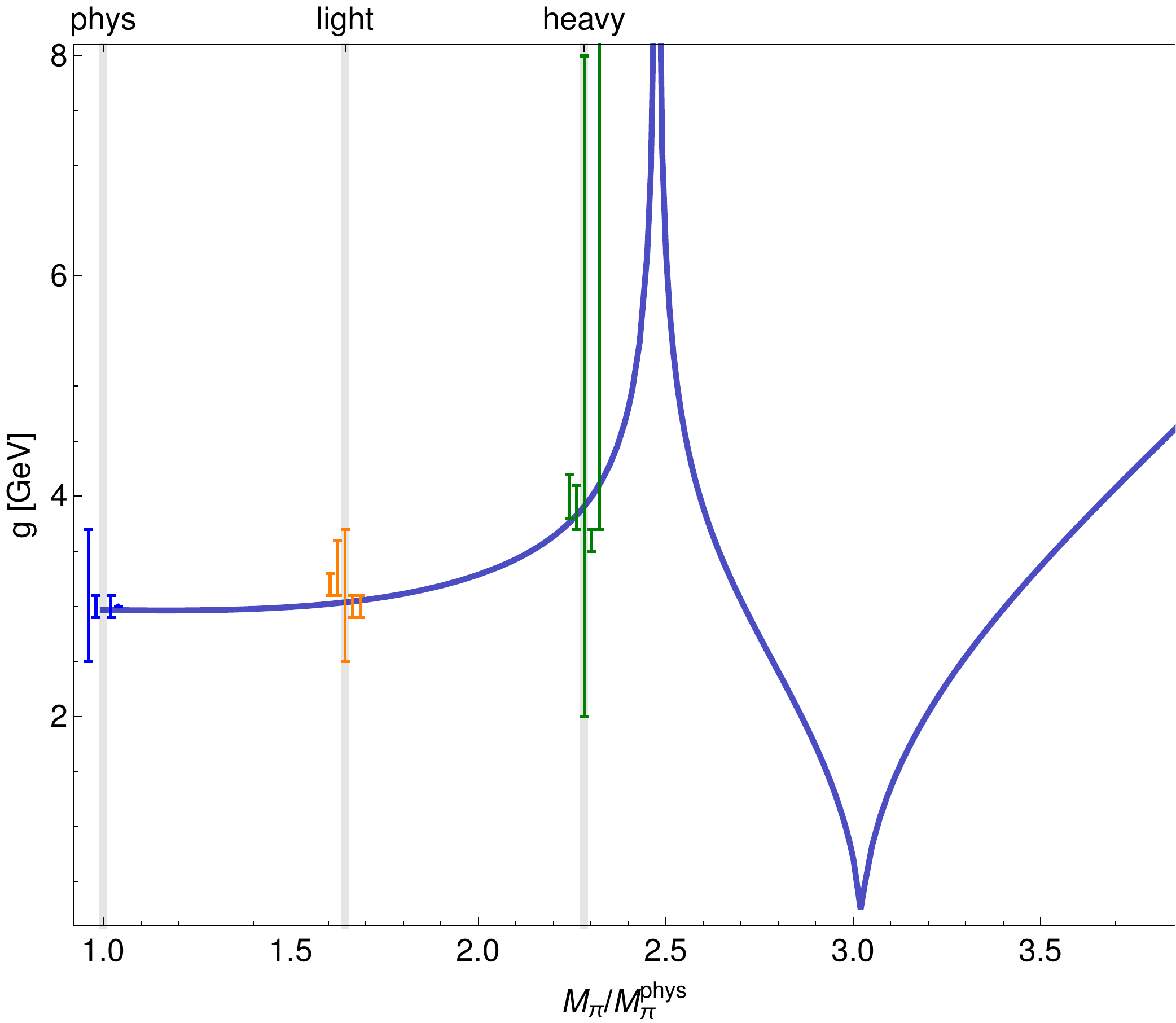}
\caption{
\label{fig:extrapolation}
Left: $M_\pi$-dependence of the pole position of the $\sigma$-resonance in the complex plane of the center of mass three-momentum from the [chm2] fit to the data from both pion masses. The dashed line represents the real   $\sqrt{s}$-axis, which connects the first ($I~RS^{\pm}$) and second ($II~RS^{\pm}$) Riemann sheets, and the subscript $+/-$ denotes the positive/negative $\sqrt{s}$ half-plane, respectively. The encircled numbers represent the pion mass in units of the physical one, while $``\mathbf{\times}"$ shows the result of the simultaneous fit to $\rho$ and $\sigma$ at light (orange) and heavy (dark green) pion masses with corresponding $1\sigma$ error areas. Right: $M_\pi$-dependence of the coupling of the $\sigma$-resonance to the $\pi\pi$ channel in the same color coding as in the left panel.
}
\end{figure*}
%%%%%%%%%%%%%%%%%%%%%%%%%%%%

With this in mind we make a prediction of the $\sigma$ pole position and the corresponding coupling to the $\pi\pi$ channel as a continuous function of the pion mass based on [chm2] fitted to both sets of lattice data, $\sigma$ and $\rho$, and both pion masses simultaneously.
To remind the reader these sets are obtained from calculations at $M_\pi\approx 1.65\,M_\pi^{\rm phys}$ and $M_\pi\approx 2.3\,M_\pi^{\rm phys}$.
The result of the extrapolation is depicted in Fig.~\ref{fig:extrapolation}.
With increasing pion mass, both poles for $\mathrm{Re}\,p$ positive and negative travel in the 2on Riemann sheet towards the $\pi\pi$ threshold, coupling more strongly to this channel. For $M_\pi\approx 2.5\,M_\pi^{\text{phys}}$, both poles meet at the real energy axis below threshold on the second Riemann sheet becoming virtual bound states. 
For higher pion mass, the poles evolve on the real axis towards and away from the $\pi\pi$ threshold ($p=0$), respectively. One of the pole reaches the two-pion threshold at $M_\pi\approx 3 M_\pi^{\text{phys}}$, where the coupling $g$ vanishes, and becomes a bound state for heavier pion masses, then, the coupling to two pions increases in this region monotonically.   
This behavior is remarkably similar to the one of Ref.~\cite{Hanhart:2008mx}.

%%%%%%%%%%%%%%%%%%%%%%%%%%%%%%%%%%%%%%%%%%%%%%%%%%%%%%%%%%%%%%%
\section{Conclusions}

To extract the parameters of the $\sigma$-resonance, we have to use a parameterization that satisfies physical constraints, in particular unitarity, analyticity, and proper chiral behavior. To gauge the systematics associated with the choice of such parametrizations, we used two types of approaches (each in several variants): a generic one that makes no assumption about the underlying dynamics, and a chiral perturbation theory inspired one that allows to extrapolate the resonance parameters to different (i.e., physical) pion mass. The systematic errors associated with the choice of parametrization are about 10\% for the pole position.

One of the strengths of the chiral parametrization is that it allows us to fit simultaneously both the $\sigma$ and $\rho$ channel, for both pion masses. We find that the model describes the data well and that the results extracted from the simultaneous fit to both channels agree well with the $\sigma$-channel fit results. We use the combined fit to extrapolate to the physical point and, based on the position of the pole in the complex energy plane, we find that ${M_\sigma = (440^{+10}_{-16}(50)-i\,240(20)(25))\MeV}$. Here the first error is the stochastic error and the second one is the combined systematic error discussed above.

The extrapolation to the physical point agrees with the 
experimental phase-shifts and the pole mass and width of the $\sigma$ is compatible with the result of recent analyses based on experimental data.

\bigskip

%%%%%%%%%%%%%%%%%%%%%%%%%%%%%%%%%%%%%%%%%%%%%%%%%%%%%%%%
\section*{Acknowledgments}
 R. M. acknowledges financial support from the Funda\c{c}\~ao de amparo \`a pesquisa do estado de S\~ao Paulo (FAPESP). D. G. and A. A. are supported in part by the
 National Science Foundation CAREER grant no. PHY-1151648 and by the U. S. DOE Grant No. DE-FG02-95ER40907. M. M. is thankful to the German Research Foundation (DFG) for the financial support, under the fellowship
 MA 7156/1-1, as well as to the George Washington University for the hospitality and inspiring environment. A. A. gratefully acknowledges the hospitality of the Physics Department at the 
 University of Maryland where part of this work was carried out. M. D. acknowledges support by the National Science Foundation (CAREER grant no. PHY-1452055) and by the U. S. Department of Energy, Office of Science, Office of Nuclear Physics under contract number DE-AC05-06OR23177. The computations were carried out on the GWU Colonial One computer cluster and the GWU IMPACT collaboration clusters; we are grateful for their support.

%%%%%%%%%%%%%%%%%%%%%

%%

\end{document}